# Cloud-Aware Web Service Security
## Information Hiding in Cloud Computing

Okal Christopher Otieno
Department of Information Technology
Mount Kenya University
Nairobi, Kenya

**Abstract:** This study concerns the security challenges that the people face in the usage and implementation of cloud computing. Despite its growth in the past few decades, this platform has experienced different challenges. They all arise from the concern of data safety that the nature of sharing in the cloud presents. This paper looks to identify the benefits of using a cloud computing platform and the issue of information security. The paper also reviews the concept of information hiding and its relevance to the cloud. This technique has two ways about it that impact how people use cloud computing in their organizations and even for personal implementations. First it presents the potential to circulate harmful information and files that can adversely affect the data those users upload on those platforms. It is also the basis of the strategies such as steganalysis and cryptographic storage architecture that are essential for data security.

### 1. INTRODUCTION

Cloud computing is still an evolving paradigm of information technology, but it has received much attention because of the capabilities it can assure the users [13]. In summary, this is a subscription-based service that allows people to use storage spaces on a network that enables them to keep and retrieve computing resources [10]. Primary examples of cloud computing technologies are emailing and social media platforms that allow a person to store different bits of information in a virtual internet space. The various enablers about cloud computing are, for example, access to one's information from anywhere in the world at any time just as long as they have an internet connection.

The scope of operation of a cloud computing services determines its type. Public Clouds are the most popular type that requires someone to use the internet to access a storage space [10]. These are the types that email users can access and also in other services that businesses have established online and any person can access their use. The private cloud consists of a small group of individuals who limit the access to space just to themselves [10]; it can involve a small organization that opens its storage for its workers only. The scope grows and develops to a community cloud that allows access between groups of organizations. When users make a combination of any of the above types, it becomes a hybrid cloud.

There are different challenges that associate with the platforms of cloud computing. They are most common in the Internet-based service providers. These internet service models have the most widespread use of the cloud [13]. Due to the higher number of persons that use the internet as a virtual storage space, there are many different risks that come up. Most of them are about managing space while ensuring the safety and efficiency of every user's operations.

These challenges open up to the research that is the primary concern of this paper. Security is one of the major problems that cloud users face around the globe [16]. It is mainly the reason that most people fear using this platform despite the benefits it promises for different enterprises. Most would prefer to meet the high costs and develop their clouds instead of taking the advantages of the cheaper Internet-based infrastructure. The author will focus on the causes of insecurity in the web-based cloud computing services. It will then outline the practices that lead to a web-aware security strategy for the providers and users that can ensure they get the best experiences. This research will focus on information hiding as a safety issue for Internet-based cloud computing applications. The researcher will take a look at various techniques of hiding data such as steganography and cryptography and how they can be dangerous or significant. The significance of different methods of ensuring that clients can experience minimal damage from these techniques is also another important consideration.

### 2. LITERATURE REVIEW

*2.1 Definition of Web-Based Cloud Computing*

There is always a common misconception that cloud computing has its scope only on the internet. It is more of a network-based strategy that enables people to share computing resources and other information without the need of setting them up on their computers [4]. These settings allow the people to perform computations on this network using standard infrastructure like software and storage spaces that are not on their computers but rather in a central location for that connection.

Web-based cloud computing provides the capability to host files and resources on an internet platform and allows the users to implement all their activities online [20]. It allows people to access a pool of these resources that they can share their connections. There are various characteristics that define web-based cloud computing [6]; first is the access to the services on demand at anytime from anywhere in the world with just an internet connection. This ability equally allows people to load the information into the system independent of their location. The cloud also provides rapid elasticity from the natures of its systems to adapt and handle the shift in workloads and storage capacity requirements from the users [9]. Another feature of these services is their measurability; it provides an easy method to estimate the value that the implementation of this strategy brings to an organization [18].

### 2.2 Benefits of Cloud Computing

There are different benefits of implementing a cloud for an organization or even an individual. Lower costs of infrastructure and operation are among the advantage that one will witness through the use of this platform [15]. The group does not need to purchase very robust infrastructure and computers to provide the ability to handle their resources [15]. There is the need to purchase on the central network equipment that will host all the services and software the people will use with any form of computing device they launch into it. The systems that use Internet clouds need a connection and perform all their functions in the virtual spaces through the websites [3]. All this capabilities lower the costs of operation for that organization and are a big gain. They also extend their scope to reduce costs in software operation and maintenance costs as the providers of the services will handle them as part of their activities [15].

Another advantage comes to a reduction in the worry about the maintenance of infrastructure and other enablers of the group's operations [14]. There is the benefit of lower dependencies on the organizational hardware and actually, they even need much less that they would if the set up their facilities [15]. This requirement reduces the probability of too much maintenance requirements of the hardware and software in the organization.

There are always issues that organizations have to handle about software licensing and updates every time. The cloud reduces the need for such requirements in an organization and defines another benefit of its implementation [12]. The service providers have to ensure that all the resources in their scope receive regular updates by allowing the internet based operations to input these changes into the system [15]. Therefore, the cloud serves the advantage that people will always use the latest version of the software if the implement the web-based services.

Using the cloud assures the clients that they have access to limitless capabilities in computing power and storage capacity [15]. The power of the entire cloud is at the disposal of the user as all they can share all the functions and abilities that all the other people on the platform can access. The need for larger storage capacities is always growing, and the providers are always expanding their facilities to accommodate the demands of their customers [15]. The segment of performance also mirrors this trend and, therefore, every user benefits from these capabilities as the access is standard to those who require high computing and storage capacities and those who do not. The clients can access much more capacities than they have on their smaller computers if they can access the cloud [15].

Data safety is another assurance that cloud computing provides to those who implement it in their operations [5]. There are different challenges that using personal computers can bring to the users. For example, there would be hardware breakdowns that endanger the data that people store in them [15]. Using the cloud eliminates these risks because the providers take care of the information in their system and can afford to conduct regular backups to secure it [5].

Compatibility is another advantage that cloud computing allows for its clients [15]. The issue of compatibility arises in the areas of operating systems, applications and document formats [15]. For example, there is no way on regular computing that an application for Windows operating systems can work in Mac personal computer. This issue does not exist in the cloud as the customers can use all the apps on any computer operating systems [15]. There is a duplication of this capability to all document formats that observe the OS boundaries and, therefore, it increases the amount of computing flexibility, unlike the other platforms. It also enables the universality of access to documents and applications which is another advantage that the users can get benefits [15].

### 2.3 Security as a Threat to Web-Based Cloud Computing

Despite the increase in popularity of cloud computing, there are different challenges in its way and one of them is data safety [25]. The use of the cloud involves complex structures of databases, networks, operating systems and the control of various other infrastructures that are mostly vulnerable [7]. The security of the information that clients put on the cloud and the different support system that enable its operations has created a great hindrance to the progress of this avenue [22]. The virtualization paradigm opens the Cloud to various problems especially in the processes of mapping the virtual space to the physical machines on the user's end [21]. There are also other challenges such as inconsistencies that are prone to arise from the processes of resource allocation and management [1]. The process also extends to the division of roles and responsibilities between the people in the organization that can lead to the same challenges [24]. These strategies determine how different users access and utilize the amenities of that cloud system. Most users will avail very sensitive information on the cloud that raises their concerns for its safety in that environment [1]. These concerns are a significant cause for lower client turnouts on the internet platforms. Since their access is public it is equally more vulnerable [8]. Security is the biggest source of worry even for the providers as it relates directly to the way that customers choose their services against competing strategies.

*2.4 Information Hiding in Cloud Computing*

Different organizations have developed various methodologies that can help reduce the impacts of the security threats. Information hiding has grown its significance in the computing arena over the recent periods [19]. It is a traditional programming method that developers use to segregate the components of a program into smaller modules to protect the whole system from the operations of a single segment [17]. There are different applications of this technique in other areas of computing such as in communication. This method is both a challenge and a savior to the security problems that different people experience in the scope of cloud computing.

Among the methods of information, hiding is steganography. It involves the practice of hiding one file behind another one of a different format to conceal the former from any monitoring in the channel of transfer [23]. The aim of this strategy is to hide a message or other data behind another probably harmless event to conceal it from all other people except for the one the sender intends it for [23]. Malicious people can use steganographic techniques to implement attacks on different computing systems or to necessitate communications that have an ill intention [13]. The most popular file formats that people use to conceal other files are graphics such as pictures and videos [13].

Regarding the steganographic procedures that people employ on different communication platforms, there is the probability of use in the cloud computing platform. It also raises the concern on how threat this cunning techniques can cause on the users of the cloud and by what scale of impact. There have been different applications of steganography in different areas to create and spread harmful files to computer systems. It has also been an efficient technique for tracking the use of copyright materials and their violations in different markets [13]. Therefore, it presents both a challenge and a way to handle some security issues in cloud computing.

The second technique for data hiding is cryptography that is a way of establishing secret codes of writing for communication between two parties to an agreement [13]. It differs from steganography in that, involves a secret for of writing whereas, steganography refers to the cover-up of different forms of information in other standard ones [13]. Therefore, one can make a cryptographic message and use the steganographic methodology to circulate it in the cloud and prevent third parties from accessing it. Cryptography also presents a dual face in the form of a problem and solution to the cloud infrastructure.

Steganography and Cryptography are the most popular forms that people sue to prevent third parties from accessing the information that their communication contains. These techniques can also be necessary for the service providers in securing the information that their clients load onto their systems in different manners of application. Most customers use these techniques to protect their data from malicious cloud users by hiding it in the storage facilities [11]. Although most people can point that this is dangerous, the providers have different ways of ensuring that their customers do not misuse this privilege to circulate damaging data in their systems.

*2.5 Protection Procedures Using Information Hiding*

Steganalysis is one of the techniques that providers use to detect if a file that a user uploads to the system has any form of hidden message within it [13]. This method does raise issues on the providers' commitment to the privacy and confidentiality of the customer and their information on the cloud [11]. But most providers implement strategies that do not violate this provisions. There are different regulatory policies that they use to develop these strategies and also ensure that the data stays in protection after analysis [11]. These techniques are in line with those of cryptographic storage services.

Steganalysis is the detection method that involves the system looking at files that have outlying characteristics such as vast sizes [2]. The primary reason for conducting this exercise is to identify the files that have steganographic codes within their structures and rendering them incapable of performing any unauthorized activity [2]. The process then proceeds to analyze the fingerprints of each file to determine the best way to handle it in the system [23]. It is not easy to develop a service that is stringent enough to deal with the shifting patterns that the cloud experiences; therefore, there is always the possibility of a threat from this forms of information hiding.

Cryptographic storage services are another technique that cloud providers use to secure the information on the facilities [11]. This method involves regular checks on the information that the users send to the cloud storage to determine if any person has tampered with it. It has an architecture that enables the system to detect and record the nature of the data that each client uploads into the cloud [11]. This section allows the cloud providers' systems to analyze and determine if the files that the customer generates are safe and appropriate to the system. When the client accesses that data in the future, the system will provide an overview of any changes or access that other people made to it during its period of existence in their storage space [11]. The systems will notify the user of how the other people that they share resources with accessed their files and inform them to reverse any undesirable changes that they made for their backups. It also enables the providers to detect any malicious activities on their system by retrieving feedback from their customers.

There are different difficulties that providers and their systems encounter in outlining these methodologies. They depend on techniques that analyze outlying file attributes that may not be unique to those that users encrypt [2]. It also takes a greater challenge in assessing the records as some of them may have irrelevant data or noise in them [2]. It also takes further effort to enable that they can decrypt the file to retrieve the information in it. It is hard to establish as standard decoding strategy; therefore, each data presents a new challenge and an increase in the workload [2]. The current methods just provide a stream of suspect files without the information on what they contain.

3. CONCLUSION

The biggest threat to cloud computing that this paper identifies is security. Most organizations worry about the safety of the information that they send and share on the cloud, especially ones that are sensitive to certain operations. There are different benefits that people can get from implementing the cloud by the threat of their information is a shadow to all of them. This trend has motivated the development of various methods and practices that can ensure the minimal possibility of such breaches.

Information hiding is both a problem and solution to the security issues in cloud computing. People can use this technique to launch attacks on customer data, and this has enormous implications that hinder the popularity of the cloud. The methods of steganalysis and cryptographic storage systems present solution to these problems as they are capable of detecting any files that contain hidden information. The only challenge remaining is to establish a universal algorithm that can detect and decrypt the information that the files contain.